\documentclass[twocolumn,showpacs,preprintnumbers,aps,prb]{revtex4}
\usepackage[dvips]{graphicx}
\usepackage{subfigure}

\begin{document}

\author{L. M. Le\'on Hilario}
\author{A. A. Aligia}
\title{$hc/4e$ oscillations in a model for (100)/(110) SQUIDs of d-wave superconductors}


\date{\today}

\begin{abstract}
We use a model of hard-core bosons to describe a SQUID built with
two crystals of $d_{x^2-y^2}$-superconductors with orientations
(100) and (110). Across the two faceted (100)/(110) interfaces,
the structure of the superconducting order parameter leads to an
alternating sign of the local Josephson coupling, and the
possibility of quartet formation. Using a mapping of the boson
model to an $XXZ$ model, we calculate numerically the energy of
the system as a function of the applied magnetic flux, finding
signals of $hc/4e$ oscillations in a certain region of parameters.
This region has a large overlap to that at which binding of bosons
exists.

\end{abstract}

\affiliation{Centro At\'omico Bariloche and Instituto Balseiro,
Comisi\'on Nacional de Energ\'ia At\'omica, 8400 S. C. de
Bariloche, Argentina}

\maketitle

\section{Introduction}
The interference between two superconducting condensates results
in a wide range of interesting phenomena. In particular, it has
been suggested that in the cuprates, the Josephson effect provides
a direct probe of the angular dependence of the superconducting
order parameter [1], and several type of phase sensitive
experiments have been performed, which confirm the
$d_{x^2-y^2}$-symmetry of the superconducting state in
high-$T_{c}$ cuprates [2].

If two crystals of cuprate superconductors are connected at an
interface, which is perpendicular to the (100) or (010) direction
in both crystals, one can have a conventional Josephson junction
(0-junction) or one with a sign reversal of the Josephson coupling
(so called $\pi$-junction) depending on the mutual orientations of
the $d_{x^2-y^2}$-wave superconducting order parameter [1] and
[3]. In principle (neglecting multiple Andreev reflections [4]),
for a perfectly flat (100)/(110) interface between two crystals of
cuprate superconductors, the CuO2 lattices meet at $45^{\circ}$,
such that a lobe (antinodal direction) of the $d_{x^2-y^2}$-order
parameter of one superconductor points towards a nodal direction
of the other, and therefore the Josephson current vanishes by
symmetry. However, microscopic roughness allows for local
Josephson tunneling across the interface facets, with the sign of
the coupling depending on the orientation of each facet [5]. This
leads to a variety of interesting effects like spontaneous
supercurrent loops [5], locally time-reversal symmetry breaking
phases [6], [7] and [8], or anomalous field dependencies of the
critical current density [9].

A particularly interesting experimental observation in
superconducting quantum interference devices (SQUIDs) with
(100)/(110) interfaces is the $hc/4e$ periodicity of the critical
current with applied magnetic flux [10]. This corresponds to
periodicity of half a flux quantum. Since the natural explanation
for the usual periodicity of one flux quantum $\Phi_{0}= hc/2e$ is
that the object that tunnels (a Cooper pair) has charge $2e$, a
possible explanation of the periodicity of $hc/4e$ is that
electrons tunnel in quartets, with total charge $4e$. Although at
first sight this possibility seems exotic, it has been proposed
before in nuclear physics, where a formation of a four-particle
condensate was predicted at low density [11]. Furthermore, it has
been proposed that pairing-quartetting competition is expected to
be a general feature of interacting fermion systems [11].
Motivated by the dominance of the second harmonics (periodicity of
$hc/4e$) of the dependence of the current with flux, Hlubina et
al. [12] studied a model for fluctuations of the phase in an array
of Josephson junctions. They conclude that the cuprates are close
to an exotic phase with quartet condensation. Furthermore, while
in principle one might think that a more conventional explanation
of half flux periodicity is the vanishing of the first harmonic in
a phenomenological Hamiltonian for Josephson junctions [10], the
work of Hlubina et al. suggests that this is related with quartet
formation. Thus, the microscopic origin of the dominance of the
second harmonic seems to be the quartet condensation.

Recently, the alternating sequence of superconducting 0- or
$\pi$-junctions (corresponding to a faceted (100)/(110) interface)
was modeled by a bosonic lattice Hamiltonian with a staggered sign
for the hopping amplitude across the facets [13]. Each boson
represents a Cooper pair and the boson hopping through the facet
is the Josephson coupling energy. As a consequence of partial
frustration of the kinetic energy, the tendency towards boson pair
formation in the presence of a weak attractive interaction is
strongly enhanced and in some region of parameters quartet
formation is favored at the interface [13]. As a consequence
$hc/4e$ oscillations are expected in closed loop with (100)/(110)
interfaces, such as the SQUIDs of the experiments of Schneider et
al. [10]. The source of attraction between pairs might be of the
same origin as the magnetic mechanism believed to be the source of
binding in the cuprates. In fact simple arguments used to justify
attraction of holes in an antiferromagnet, or a system with short
range antiferromagnetic order, can be easily extended to more
particles [13]. However, usually the loss of kinetic energy does
not allow binding of more than two particles.

Since Josephson-junction-type Hamiltonians can be derived from a
boson lattice model when fluctuations of the amplitude of the
superconducting order parameter are integrated out [14] and [15],
as stated above, one might expect that the approaches of Refs.
[12] and [13] are related and that the appearance of higher
harmonics in the Josephson current is directly connected with
quartet formation.

In this work, we calculate the energy as a function of the flux
$E(\Phi)$ for a simple model describing the motion of Cooper pairs
in a SQUID with two (100)/(110) interfaces, as in the experimental
situation [10] (see Fig. 1). Since the current through the ring
$j(\Phi)$ is proportional to $\partial E/\partial \varphi$ for a
given geometry, the oscillations in $E(\Phi)$ are directly related
to those of the current. While some ideas were reported before
[13], the calculation of pairing was limited to the dilute case of
very few bosons in the system, and $E(\Phi)$ was not calculated.
In Section 2 we explain the model used and its mapping to a spin
1/2 model. The results of the numerical solution of this model are
presented in Section 3. Section 4 contains a short summary and
discussion.
\begin{figure}[htbp]
    \includegraphics[width=1.0\linewidth]{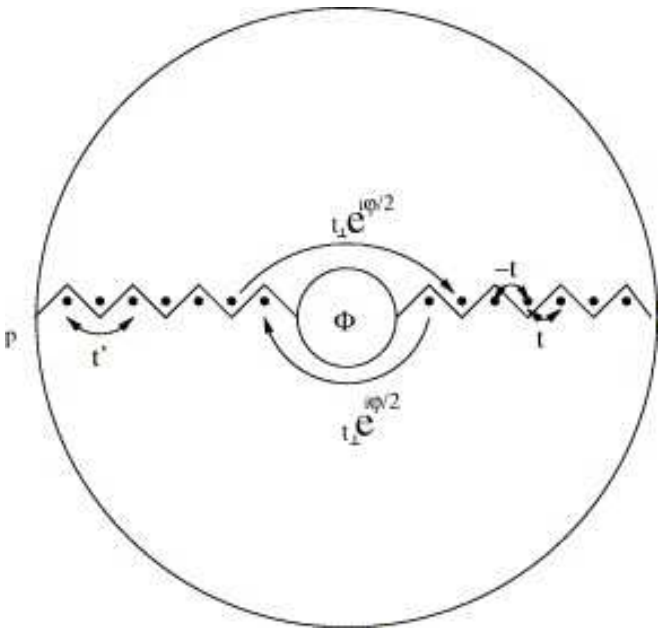}
    \caption{Sketch of a SQUID with two faceted (100)/(110)
    interfaces and the simple Hamiltonian Eq. (1) used to describe it.}
    \label{rrs1}
\end{figure}

\section{The Model}
We describe the Cooper pairs as bosons moving on a lattice [13].
At the interfaces, the boson hopping across a facet represents the
Josephson coupling energy of the quantum phase Hamiltonian.

The Hamiltonian is [13] (see Fig. 1)
\begin{eqnarray}
H&=&H_{1} + H_{2} + H_{3}\nonumber \\
H_{1}&=&\sum_{\alpha i}(-t(-1)^{i}a_{\alpha
i+1}^{\dagger}a_{\alpha
i}- t'a_{\alpha i+2}^{\dagger}a_{\alpha i}+H.c.), \nonumber \\
H_{2}&=&\sum_{\alpha i}[V a_{\alpha i}^{\dagger}a_{\alpha
i}a_{\alpha i+1}^{\dagger}a_{\alpha i+1}+U(a_{\alpha
i}^{\dagger}a_{\alpha i}-1)a_{\alpha i}^{\dagger}a_{\alpha
i}],\nonumber\\
H_{3}&=&-t_{\perp}\sum_{j}(a_{1j}^{\dagger}a_{2j}e^{i\varphi(-1)^{j/2}}+H.c.)
\end{eqnarray}
where $a_{\alpha i}^{\dagger}$ creates a boson (Cooper pair) at
site $i$ of the interface $\alpha$= 1,2. The first term $H_{1}$
describes the kinetic energy of the bosons within each interface.
The hopping with amplitude t describes Josephson tunneling across
the (100)/(110) interface and has an alternating sign due to the
roughness of the surface and the different change of phase of the
superconducting order parameter. The term in $t'$ describes motion
along the interface without crossing it. The second term $H_{2}$
represents an on-site $U$ and a nearest-neighbor $V$ interaction.
The last term $H_{3}$ describes in the simplest possible way the
motion of Cooper pairs from one interface to the other. The phase
$\varphi$ is related to the magnetic flux threading the ring by
$\varphi = 2\pi \Phi/\Phi_{0}$, where $\Phi_{0} = hc/2e$ is the
flux quantum. For simplicity we take $U \rightarrow +\infty$. This
implies hard-core bosons and allows to map the model into a
spin-1/2 model as described below. The nearest-neighbor
interaction is assumed attractive ($V < 0$) unless otherwise
stated. Its origin can be the same as the magnetic pairing
interaction in the cuprates [13]. From results of the equivalent
spin-1/2 $XXZ$ model for one interface [16], we expect that
inclusion of second nearest-neighbor attraction does not change
essentially the physics. The values of the parameters were
estimated as $V\sim -250$ K, $t' > t\sim  100$ K [13]. These
parameters favor quartet formation at one (100)/(110) interface,
but not in conventional interfaces with 0-junctions only [13].

The alternating sign in the first term of Eq. (1) can be
eliminated through an adequate change of phases of the boson
operators,
\begin{eqnarray}
b_{\alpha4i}^{\dagger}=-a_{\alpha4i}^{\dagger},\;
b_{\alpha4i+1}^{\dagger}=-a_{\alpha
i+1}^{\dagger},\nonumber\\
b_{\alpha4i+2}^{\dagger}=a_{\alpha4i+2}^{\dagger},\;
b_{\alpha4i+3}^{\dagger}=a_{\alpha4i+3}^{\dagger}
\end{eqnarray}
which transforms the first term into
\begin{equation}
H_{1}=\sum_{\alpha i}[-tb_{\alpha i+1}^{\dagger}b_{\alpha i} +
t'b_{\alpha i+2}^{\dagger}b_{\alpha i}+H.c.]
\end{equation}
while the other terms retain the same form with the replacement
$a_{\alpha i}\rightarrow b_{\alpha i}$.

Using a boson-spin transformation
\begin{equation}
S_{\alpha i}^{+}=(-1)^{i}b_{\alpha i}^{\dagger},\;S_{\alpha
i}^{-}=(-1)^{i}b_{\alpha i},\;S_{\alpha i}^{z}=b_{\alpha
i}^{\dagger}b_{\alpha i}-\frac{1}{2}
\end{equation}
the model is mapped into the following spin-1/2 model
\begin{eqnarray}
H_{S}&=&\sum_{\alpha i}[J_{1}(S_{\alpha i}^{x}S_{\alpha
i+1}^{x}+S_{\alpha i}^{y}S_{\alpha i+1}^{y}+\Delta S_{\alpha
i}^{z}S_{\alpha i+1}^{z})\nonumber\\
&+&J_{2}(S_{\alpha i}^{x}S_{\alpha i+2}^{x}+S_{\alpha
i}^{y}S_{\alpha
i+2}^{y})]\nonumber\\
&+&J_{3}\sum_{j}(e^{i\varphi(-1)^{j/2}}S_{1j}^{+}S_{2j}^{-}+H.c.)/2
\end{eqnarray}
with the spin exchange coupling constants $J_{1} = 2t, J_{2} =
2t', J_{3} = 2t$, and the anisotropy parameter $\Delta = V/2t <
0$. For $J_{3} = 0$, the model describes two uncoupled $XXZ$
chains representing the interfaces. If periodic or antiperiodic
boundary conditions are used for each chain, the Hamiltonian Eq.
(5) can be thought as describing two $XXZ$ rings, one on top of
the other with a complex spin–flip interaction $J_{3}$ between
them, the phase of which alternates sign between odd and even
sites of the rings. The spin-1/2 $XXZ$ chain has been studied
before in the context of metamagnetic transitions and results for
magnon binding (bosons in the original language) and phase
separation were obtained in specific parameter regimes [16]. The
region of parameters for which binding of two bosons occur, and
clustering in more than two bosons or phase separation takes place
has been studied in more detail in Ref. [13]. Taking $t = 1$ as
the unit of energy, the region most favorable for pairing of
bosons can be defined roughly as $t' > 1$ and $1 < -V < 2t' + 2$.
In the following section, we report our search for signals of
$hc/4e$ oscillations in $E(\Phi)$ in parameter space and for a
finite concentration of bosons.

For this model, the current in the loop is given by [17]
\begin{equation}
j(\Phi)=\frac{2e}{\hbar}\frac{\partial \langle H\rangle}{\partial
\varphi}=\frac{2eN_{s}}{\hbar}\frac{\partial E(\Phi)}{\partial
\varphi}
\end{equation}
where $N_{s} = 2L$ is the number of sites in the system, and $L$
is the number of sites in one ring.

\section{Results}

We have studied by numerical diagonalization the equivalent spin
Hamiltonian Eq. (5) for two rings of $L$ sites each and a number
$N$ of up spins (bosons in the original language) in a background
of down spins. The total spin projection $S_{z} = N - L/2$ is a
conserved quantity in Eq. (5). In general, we have taken $8 \leq L
\leq 12$. For small $N$ some calculations were done also for $L =
14$ (28 sites counting both rings). We have chosen for each ring
the boundary conditions (BC) periodic or antiperiodic, which lead
to the minimum ground state energy per site as a function of flux
$E(\Phi)$. In all studied cases, these BC did not change as the
flux $\Phi$ was varied. Actually in general, and particularly in
the region of binding, the BC which lead to the minimum energy are
twisted [16] and [19]. However, the deviation from periodic or
antiperiodic BC is small for $t'\geq  t$ and we believe that our
choice does not affect the results.

The energy per site $E(\Phi)$ is periodic in one flux quantum:
$E(\Phi) = E(\Phi + \Phi_{0})$. We have investigated the tendency
towards a periodicity of half a flux quantum in $E(\Phi)$. This is
reflected in the presence of two relative minima in $E(\Phi)$ for
$\Phi = 0$ and $\Phi = \Phi_{0}/2$. For $N = 2$ and $N = 4$, we
have explored the region of parameters $-3 \leq t'/t \leq 3, -8
\leq V/t \leq 0$, examining in more detail the region of binding
determined previously [13] (roughly $t'/t > 1$ and $1 < -V/t < 2t'
+ 2, t > 0$). For only two bosons in the system ($N = 2$), the
region of binding has been determined analytically in the
thermodynamic limit [13] and [16]. Rather surprisingly, we do not
find two minima for $N = 2$ in all the explored region of
parameters. Instead, for $N = 4$, we do find clear signals of a
periodicity in $\Phi_{0}/2$, and only inside the region of
binding. An example is shown in Fig. 2 for $t'/t > 1$ and $L = 8$.
Similar results are obtained for larger system sizes and
moderately larger values of $t'/t$. However, keeping $N = 4$ and
increasing $L$ up to $L = 14$, the relative minimum of $E(\Phi)$
at $\Phi = 0$ becomes less pronounced, suggesting that it
disappears and there is no periodicity of $E(\Phi)$ in
$\Phi_{0}/2$ in the thermodynamic limit for dilute systems $N/L
\rightarrow 0$. In any case, the dilute limit is not expected to
be relevant to the experimental situation.
\begin{figure}[htbp]
    \includegraphics[width=1.0\linewidth]{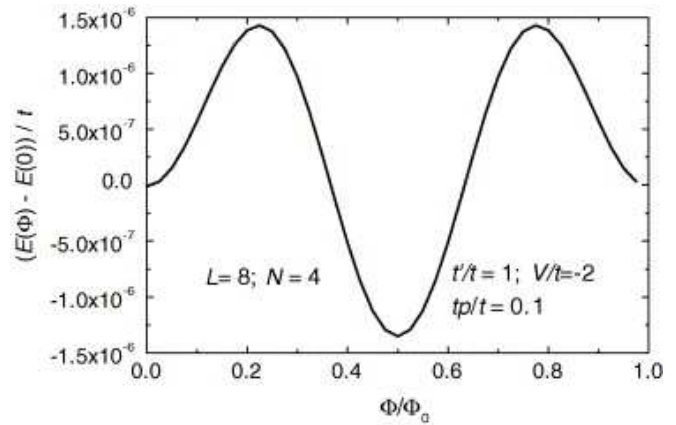}
    \caption{Ground state energy per site as a function
    of the applied flux for $L = 8. N = 4, t'/t = 1.5, V/t =
-2$ and $t_{\perp}/t = 0.1.$}
    \label{rrs2}
\end{figure}
To be able to analyze how $E(\Phi)$ evolves with system size at a
finite fixed concentration of bosons, we calculated $E(\Phi)$ for
increasing $L$, keeping $N = L$ and the other parameters fixed.
Note that due to the symmetry of the spin Hamiltonian Eq. (5)
under a change of sign $S_{z}$, the problems with $N$ or $2L - N$
bosons are equivalent. Therefore, the case with $N = L$ is that of
the greatest possible total kinetic energy in absence of the
interaction $V$. In Fig. 3 we show how $E(\Phi)$ changes as $L$
increases from 8 to 12 (in even steps to avoid frustration) and
other parameters as in Fig. 2. The absolute minimum is at $\Phi =
0$ and there is a relative minimum at $\Phi = \Phi_{0}/2$. This
minimum is rather shallow for $L = 8$, but becomes more pronounced
as $L$ increases. This is consistent with a periodicity of
$E(\Phi)$ in half a flux quantum in the thermodynamic limit.
\begin{figure}[htbp]
    \includegraphics[width=1.0\linewidth]{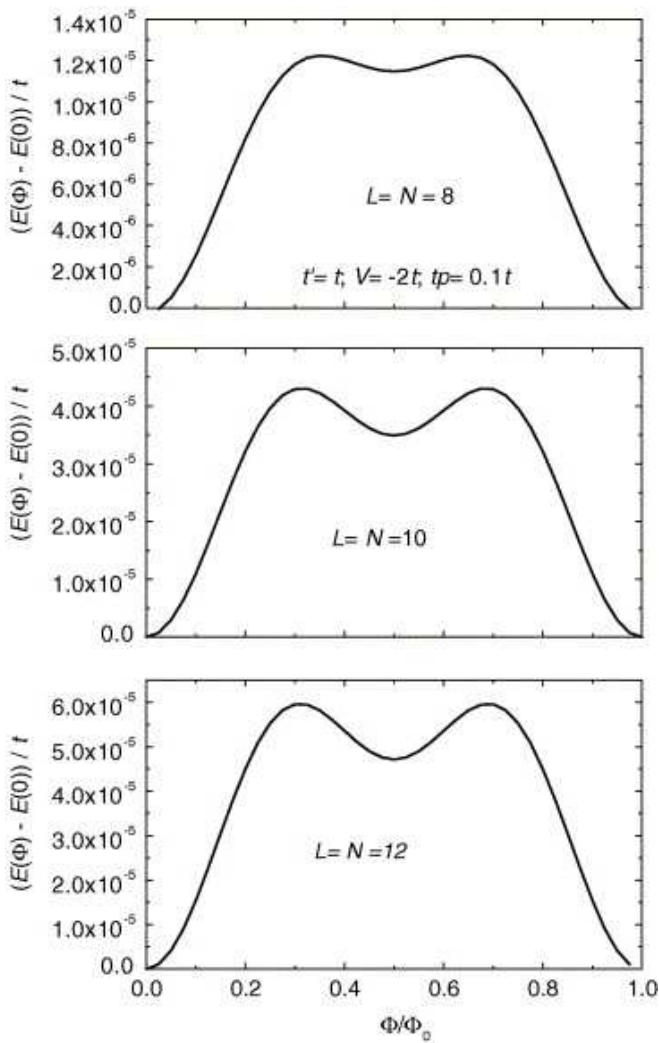}
    \caption{ Ground state energy per site as a function of the applied flux
    for $t'/t = 1, V/t = -2, t_{\perp}/t = 0.1, N = L$ and different system sizes $L$.}
    \label{rrs3}
\end{figure}
For a reasonable value of the Josephson coupling energy $t \sim
10^{-2}$ eV, one has $4et/\hbar \simeq 10 \mu A$. Since the number
of facets in a real system is of the order of $L = 10^{3}-10^{4}$
[18], from Eq. (6), the current extrapolated to the thermodynamic
limit is $j \sim 1-10 \mu A$. This in rough agreement with the
experimental order of magnitude $j \sim 10-100 \mu A$ [10]. Since
roughly $j \sim t_{\perp}^{2}$(see below) the agreement improves
if $t_{\perp}$ is increased.

The evolution of $E(\Phi)$ with size for larger $t'$ is shown in
Fig. 4. In the spin language, these parameters correspond to a
next-nearest-neighbor antiferromagnetic interaction in the $x$ and
$y$ directions $J_{2}$ two times larger as the corresponding one
for nearest neighbors $J_{1}$. In this situation, in the classical
case, an antiferromagnetic order between next-nearest neighbors is
favored and to avoid frustration of this order in a ring, the
number of sites L should be multiple of 4. As a consequence,
although this frustration is only partial in the quantum case, we
believe that the results for $N = L = 10$ are not reliable for $t'
> t$. This is the case of the middle panel in Fig. 4. Comparing the
other two cases represented in Fig. 4, although none of them shows
two well defined minima, the results are not inconsistent with a
double periodicity in the thermodynamic limit because $E(\Phi)$
has a rather sharp maximum for $N = L = 8$ at $\Phi = 0.5\Phi$
that becomes rather flat for $N = L = 12$, suggesting the
development of a minimum as $L$ is further increased. Note that
for $t_{\perp}/t = 0.2$, a relative minimum in $E(\Phi)$ at $\Phi
= 0.5\Phi_{0}$ already exists for $N = L = 12$ (dashed line in
Fig. 4).
\begin{figure}[htbp]
    \includegraphics[width=1.0\linewidth]{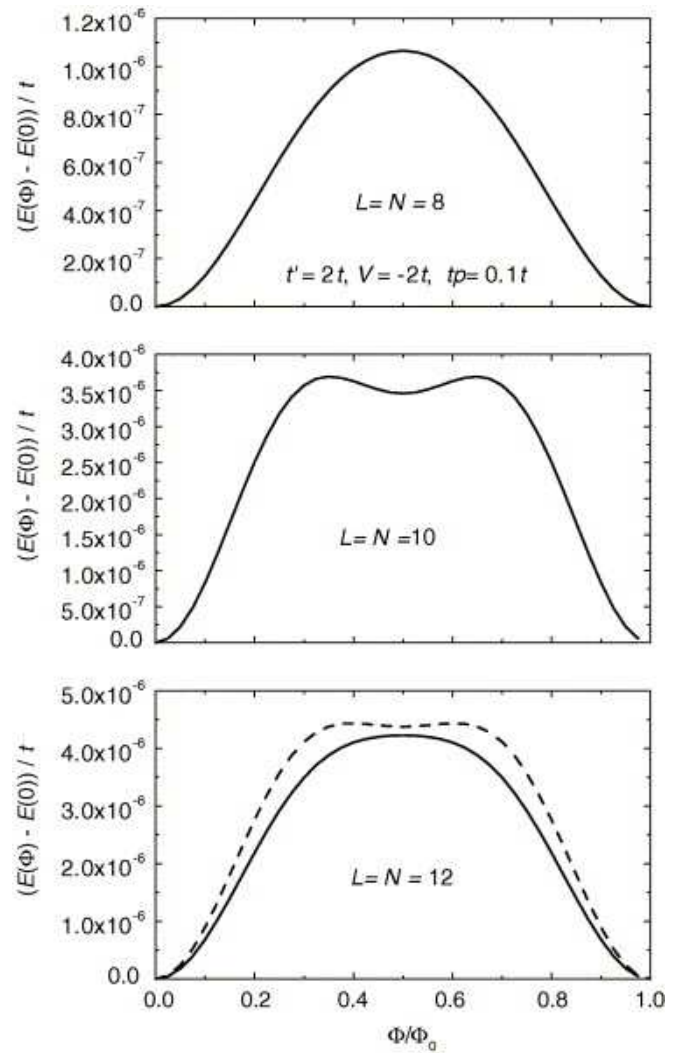}
    \caption{Same as Fig. 3 for $t'/t = 2$. The dashed line at the bottom
    is the result for $t_{\perp}/t = 0.2$ multiplied by a factor 0.2. }
    \label{rrs4}
\end{figure}
We have also analyzed the effect of varying hopping between
interfaces $t_{\perp}$. As $t_{\perp}$ increases, the tendency
towards an additional periodicity increases slightly up
$t_{\perp}/t \sim 0.2$ and for larger $t_{\perp}$, it is weakened.
This is expected, since the additional kinetic energy tends to
break the bound pairs. For small $t_{\perp}$, from perturbation
theory, the structure of $E(\Phi)$ is dominated by several terms
of order $t_{\perp}^{2}$ . This is in rough agreement with our
numerical results for $t_{\perp}/t \sim 0.2$, although the
increase in the amplitude of $E(\Phi)$ with $t_{\perp}$ seems
faster than quadratic (see for example the bottom of Fig. 4).

Most of the above-mentioned results correspond to the half filled
case $N = L$. We have also looked for signals of an extra
periodicity in $E(\Phi)$ in the quarter filled case $N = L/2$,
comparing the flux dependence of the energy for $L = 8$ and $L =
12$. The results suggest that no double periodicity is present in
the thermodynamic limit.

Finally, we also study the shape of $E(\Phi)$ for repulsive
interaction $V > 0$ in the half filled case. Surprisingly, we find
a tendency for double periodicity for positive $t'$ (see Fig. 5).
We have studied the binding energy of an isolated ring $\Delta_{b}
= L[E_{1}(N + 2) + E_{1}(N) - 2E_{1}(N + 1)]$ for different
parameters, where $E_{1}(N)$ is the energy per site of one ring
for $N = L/2$ particles. The region of binding coincides roughly
with that obtained previously for $N = 0$ [13] and [16] and
$\Delta_{b} > 0$ for positive $V$. The absence of a negative
binding energy for positive $V$ also persist when $t_{\perp}$ is
included. The factor of 10 increase in the magnitude of the
oscillations as $L$ is increased form 8 to 12 points to
particularly large finite-size effects.
\begin{figure}[htbp]
    \includegraphics[width=1.0\linewidth]{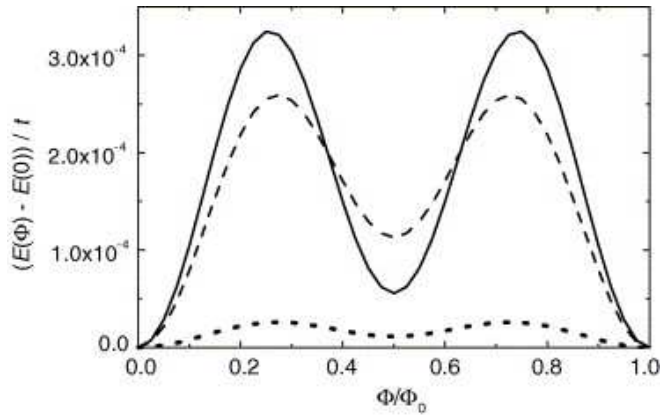}
    \caption{Ground state energy per site as a function of the applied
    flux for $t'/t = 1.5, V/t = -2, t_{\perp}/t = 0.2$ and different values of $N = L$:
    full line $L = 12$, dotted line $L = 8$, dashed line 10
times $L = 8$.}
    \label{rrs5}
\end{figure}
In all cases we do not find signs of extra periodicity for
negative $t'$. We remind the reader that negative $t'$ corresponds
to the case of ordinary junctions in the facets (all 0-junctions),
for which no extra periodicity was experimentally observed.

\section{Summary and discussion}
We have calculated numerically the energy per site as a function
of applied magnetic flux $E(\Phi)$ in a simplified model for
hard-core bosons (representing Cooper pairs) for a SQUID
containing two (100)/(110) interfaces. In spite of the limitations
of the size of the systems studied, the finite-size scaling
suggests $hc/4e$ periodicity for parameters for which binding of
bosons at one interface take place [13]. This fact and the
magnitude of the resulting current is consistent with the
experiments of Schneider et al. [10]. This periodicity is absent
for usual interfaces ($t'< 0$ in the model).

We also find signals of $hc/4e$ oscillations in $E(\Phi)$ for
repulsive interactions for which no binding exists. The reason of
the extra periodicity in this case is unclear, and needs further
study.

For simplicity the motion of Cooper pairs inside one crystal from
one interface to the other has been represented by only one
hopping parameter $t_{\perp}$. This is a rather crude description
which allowed us to include more sites at the interfaces, keeping
the total amount of sites below 32 due to computer limitations. In
any case, we expect that the essential physics is retained, at
least for attractive interaction between Cooper pairs. The
physical picture can be the following: quartets (bound pairs of
bosons) are formed at the interfaces as a consequence of the
competition between attraction and the reduced kinetic energy
there [13]. The energy scale of the motion perpendicular to the
interface ($t_{\perp}$ in our model) is responsible for the
coherence and the observed flux dependence. However, if it is too
large, it tends to unbind the quartets. If this image is correct,
and quartets live only near the interface, the observation of
$hc/4e$ oscillations should depend on the size of the system. They
should be weaker or disappear if the coherence length in the
direction normal to the interface is much smaller than the size of
the system.

A more realistic model for the description of the motion of Cooper
pairs between both interfaces should include one or more layers of
intermediate sites between the interfaces.

\section*{Acknowledgments}
A.A.A. wants to thank A. Kampf, T. Kopp and J. Mannhart for useful
discussions. This work was sponsored by PICT 03-13829 of ANPCyT.
L.M.L.H is a fellow of the IB-ICTP Diploma Program. A.A.A. is
partially supported by CONICET.

\end{document}